\documentclass[twocolumn,aps,prd,amsmath,amssymb,nofootinbib]{revtex4-1}\newcommand{\preprintnumber}{\hfill MIT-CTP 4671\,\,\,\,\maketitle}

\usepackage{graphicx,bm,enumerate}

\newcommand{\beq}{\begin{equation}}
\newcommand{\be}{\begin{equation}}
\newcommand{\bea}{\begin{eqnarray}}
\newcommand{\eeq}{\end{equation}}
\newcommand{\ee}{\end{equation}}
\newcommand{\eea}{\end{eqnarray}}

\newcommand{\mpl}{M_{\rm Pl}}
\newcommand{\lpl}{l_{Pl}}
\newcommand{\mpls}{M_{*}}
\newcommand{\muv}{m_{\mbox{\tiny{UV}}}}
\newcommand{\ex}{d}
\newcommand{\rad}{\psi}

\newcommand{\exx}{\tilde{d}}

\newcommand{\bel}[1] {\begin{equation}\label{#1}}
\newcommand{\beal}[1] {\begin{eqnarray}\label{#1}}

\def\CalV{{\cal V}}
\def\({\left(}
\def\){\right)}
\def\[{\left[}
\def\]{\right]}

\def\nn{\nonumber \\}

\begin{document}

\title{Can Compactifications Solve the Cosmological Constant Problem?}

\author{Mark P.~Hertzberg$^{1,2}$}
\author{Ali Masoumi$^1$}
\affiliation{$^1$Institute of Cosmology, Department of Physics and Astronomy,\\
Tufts University, Medford, MA 02155, USA}
\affiliation{$^2$Center for Theoretical Physics, Department of Physics,\\ Massachusetts Institute of Technology, Cambridge, MA 02139, USA}

\date{\today}

\begin{abstract}
Recently, there have been claims in the literature that the cosmological constant problem can be dynamically solved by specific compactifications of gravity from higher-dimensional toy models. These models have the novel feature that in the four-dimensional theory, the cosmological constant $\Lambda$ is much smaller than the Planck density and in fact accumulates at $\Lambda=0$. Here we show that while these are very interesting models, they do not properly address the real cosmological constant problem. As we explain, the real problem is not simply to obtain $\Lambda$ that is small in Planck units in a toy model, but to explain why $\Lambda$ is much smaller than other mass scales (and combinations of scales) in the theory. Instead, in these toy models, all other particle mass scales have been either removed or sent to zero, thus ignoring the real problem. To this end, we provide a general argument that the included moduli masses are generically of order Hubble, so sending them to zero trivially sends the cosmological constant to zero. We also show that the fundamental Planck mass is being sent to zero, and so the central problem is trivially avoided by removing high energy physics altogether. On the other hand, by including various large mass scales from particle physics with a high fundamental Planck mass, one is faced with a real problem, whose only known solution involves accidental cancellations in a landscape.
\end{abstract}

\preprintnumber


\section{Introduction}

\let\thefootnote\relax\footnotetext{Electronic address: {\tt mark.hertzberg@tufts.edu}\\Electronic address: {\tt ali@cosmos.phy.tufts.edu}}

A range of cosmological observations indicate that the current expansion rate of the universe is accelerating \cite{Riess:1998cb}. This is accommodated within general relativity by the introduction of the so called cosmological constant $\Lambda$. Viewed as a source of vacuum energy density, it has an equation of state $w =-1$. By taking its energy density to be $\sim 70\%$ of the current critical density, it leads to acceleration in a fashion that is beautifully compatible with current data \cite{Hinshaw:2012aka, Ade:2013uln}. On the one hand, this is another spectacular triumph for general relativity and particle physics, which suggests the appearance of vacuum energy. On the other hand, typical estimates for the value of the vacuum energy are many orders of magnitude larger than the observed value of $\Lambda_{\rm obs}\sim (10^{-3}\,\mbox{eV})^4$. This leads to the problem of why the observed cosmological constant is so small; for a review see Ref.~\cite{Martin:2012bt}.

Many proposals have tried to address this problem, often involving radical modifications to the structure of gravity or quantum field theory.
One class of recent proposals stays within the framework of ordinary field theory, but appeals to the existence of extra dimensions \cite{Brown:2013fba,Brown:2014sba}. An appreciable vacuum energy is included in the higher-dimensional theory and various other sources of energy; fluxes and curvature (for earlier work on flux compactifications, see \cite{Douglas:2006es,Denef:2007pq}, and for some difficulties in achieving a positive cosmological constant in string compactifications, see \cite{Hertzberg:2007ke,Hertzberg:2007wc}). It is then found that upon compactifying to 4 dimensions, the resulting lower-dimensional cosmological constant is arbitrarily small and has an accumulation point at $\Lambda=0$. This appears to be a wonderful solution to the cosmological constant problem. At the same time, it is acknowledged that this solution comes at a price; it predicts the existence of additional arbitrarily light scalars, with masses of order Hubble. 

In this paper, we show that the lightness of various scales, completely undermines the whole purported solution. 
As we explain, the heart of the cosmological constant problem is to explain why $\Lambda$ is so small despite the presence of various heavy scales (such as the Standard Model fields, and possible heavier fields associated with unification and quantum gravity). It misses the essential problem to merely send all mass scales to zero in a toy four-dimensional theory (in units of the four-dimensional Planck mass). Moreover, we show that in order for the cosmological constant to be small in these models, the fundamental Planck scale is also being sent to zero; thus removing any high energy completely. However the heart of the cosmological constant problem is to explain why $\Lambda$ is small despite the existence of high energy physics, including heavy fields and a high fundamental scale. Related to this, we clarify some issues surrounding the problem in different setups, such as electromagnetism and gravity, and provide a general explanation as to why the moduli masses are naturally of order Hubble and how $\Lambda$ is linked to the higher-dimensional fundamental scale.

Our paper is organized as follows: 
In Section \ref{Free} we describe two different notions of fine-tuning. 
In Section \ref{E&MGrav} we describe fine-tuning in two different models.
In Section \ref{Comp} we discuss a class of compactification models. 
In Section \ref{Mass} we compute the moduli masses and the fundamental Planck mass in this class of models. 
Finally, in Section \ref{Disc} we conclude.
We work in units in which $c=1$, but we will keep powers of $\hbar$ to track classical versus quantum effects. We will write the ``mass" couplings in the field theory as $m$, even though $m$ in classical field theory is actually a frequency, and the mass of quanta is $\hbar\,m$.

\section{Notions of Fine-Tuning}\label{Free}

\subsection{Sharp Cutoff Analysis}\label{Sharp}

To set the stage for our later argument, it is useful to study the vacuum energy in free theories. Of course the vacuum energy only has consequences when we include gravitation, so we mean ``free" in the particle sector.

The vacuum energy receives a quantum contribution $\Lambda_{\rm quant}$ from a one loop diagram. 
It is well known that this leads to the following vacuum energy
\beq
\Lambda_{\rm quant}= \pm \,g\,\hbar\! \int\! {d^3k\over(2\pi)^3}\,{\omega_k\over 2}
\eeq
where $g$ is the number of degrees of freedom, $+$ is for bosons, $-$ is for fermions, and $\omega_k=\sqrt{k^2+m^2}$ (we allow for massive free particles). It follows that quantum corrections provide a quartic divergence $\Lambda_{\rm quant}\sim \hbar\, m_{\mbox{\tiny{UV}}}^4$, and if we put $m_{\mbox{\tiny{UV}}}=1/\sqrt{\hbar\,G}$ (the Planck frequency) then we have a Planck energy density. 
On the other hand, the total contribution to the vacuum energy receives a contribution from the ``bare" or ``classical term" $\Lambda_{\rm bare}$, so that the total vacuum energy is
\beq
\Lambda = \Lambda_{\rm bare} + \Lambda_{\rm quant}
\eeq
Hence in order for $\Lambda$ to be much less than the cutoff density, requires an extreme {\em fine-tuning} between these two contributions. So it appears as though any model which can dynamically produce a very small $\Lambda$, especially if $\Lambda$ can be made arbitrarily small, is a solution of the cosmological constant problem. 

Moreover, if we define the momentum integral with a UV cutoff $m_{\mbox{\tiny{UV}}}$ and expand in powers of $m/m_{\mbox{\tiny{UV}}}$ we obtain
\bea
\Lambda_{\rm quant} &=& \pm\,{g\,\hbar\over(4\pi)^2}\Bigg{[}c_1\,\muv^4+c_2\,\muv^2\, m^2\nonumber\\
&&\,\,\,\,\,\,\,\,\,\,\,\,\,\,\,\,\,\,\,\,\,\,
+c_3\,m^4\ln\!\left(m^2\over\muv^2\right)+\ldots\Bigg{]}
\label{rhoexp}\eea
where $c_{1,2,3}=\mathcal{O}(1)$ numbers that depend on choice of regularization. This is the most general expansion for a field at one-loop. 
The first term provides the usual claim of a quartic contribution to vacuum energy. It suggests that even if $m=0$, there must be tremendous fine-tuning to cancel against $\Lambda_{\rm bare}$ in order for $\Lambda$ to be small.

\subsection{Renormalization Group Analysis}\label{Renormalization}

The above analysis is highly suggestive that there is a quartic sensitivity to the cutoff, requiring significant fine-tuning to avoid a large $\Lambda$. 

One could, however, focus on another notion of ``fine-tuning". To explain this, we should recall that the only physical parameters of a theory are the {\em renormalized} couplings, rather than the bare couplings or quantum corrections, which are scheme dependent. Such couplings are defined at some renormalization scale and change according to the renormalization group. By including gravity, this includes the physical cosmological constant $\Lambda$. We can write 
\beq
\Lambda=\Lambda(\mu)
\eeq
where $\mu$ is some renormalization scale. Within the framework of local quantum field theory, one physical notion of fine-tuning is that there is a delicate cancellation among renormalized parameters in order to fit the data. In standard renormalization schemes, the quartic and quadratic diverges of eq.~(\ref{rhoexp}) can be absorbed by $\Lambda_{\rm bare}$, while the logarithm, proportional to $\hbar\,m^4$, is associated with an actual flow of the coupling. Further discussion of these issues includes Refs.~\cite{Babic:2001vv,Burgess:2013ara}.

In particular, as we flow from some high scale $\mu_H\gg m$, down to some low scale $\mu_L\ll m$, there is a jump in the renormalized coupling from passing through a mass scale of the order $\sim\hbar\,m^4$. If we pass through several mass scales, denoted $m_i$, the change is roughly
\bea
\Delta\Lambda & = & \Lambda(\mu_H)-\Lambda(\mu_L)\\
& \sim & {\hbar\over(4\pi)^2}\sum_i(\pm)_i\,g_i\,m_i^4
\eea
where we suppress possible logarithmic and threshold effects.
Hence, in order for $\Lambda_{\rm obs}\approx\Lambda(\mu_L)$ to be very small, there must be some exquisite cancellation between the renormalized coupling at a high scale $\Lambda(\mu_H)$ and the sum and differences of various renormalized masses $\sim\hbar\,m_i^4$.

Conversely, if one investigates theories that are built out of massless or extremely light particles (and no dynamically generated scales due to strong dynamics) then the flow of the renormalized $\Lambda(\mu)$ will be small, and it does not require fine tuning. We will return to this issue when considering a class of compactification models.

\subsection{Summary}

In physical models, we generally study interesting theories with various heavy particles, and the challenge is to explain how the observed cosmological constant is small compared to the quartic power of some scale. From the point of view of the ``sharp cutoff analysis", we compare $\Lambda_{\rm obs}$ to $\hbar\,\muv^4$, and from the point of view of the ``renormalization group analysis", we compare $\Lambda_{\rm obs}$ to $\hbar\,m_i^4$, where $m_i$ is the heaviest mass scale. Often these two scales will not differ too much anyhow, as often heavy fields appear around the cutoff of an effective theory. If we take $\muv\sim\mpl$, or we suppose there are particles whose mass is close to $m_i\sim\mpl$, then in either point of view we expect a Planck density for $\Lambda$. Moreover, there can be additional classical field contributions to the vacuum energy in interacting theories (for example, from scalar potentials or when gravity is included) that can be large too, but this again depends on the presence of high scales. Let us now further explore this in some important interacting theories.

\section{Vacuum Energy Examples}\label{E&MGrav}

\subsection{Pure Electromagnetism}\label{E&M}

As a warm-up to the gravitational case, we start here by studying the problem of vacuum energy from photons.
We consider the following interacting theory of only massless photons minimally coupled to gravity
\beq
S = \int\! d^4x\sqrt{-g}\left[-\Lambda-{1\over4}F_{\mu\nu}F^{\mu\nu}+{1\over M^4}(F_{\mu\nu}F^{\mu\nu})^2+\ldots\right]
\label{EM}\eeq
As shown in the previous section, introducing a sharp cutoff on the vacuum energy loop integral reveals a vacuum energy contribution $\Lambda_{\rm quant}\sim\hbar\, \muv^4$. However if we focus on the renormalization group flow in a local Lorentz invariant theory, we see that there is no flow from massless fields, and the higher order interaction term does not change this conclusion (at least for energies well below the cutoff). Hence a theory of massless photons does not strictly speaking have any cosmological constant problem from the point of renormalization group flow.

However the interaction term renders the theory non-renormalizable. So one expects new physics to enter at some scale $m_{\rm new}$ satisfying $m_{\rm new}<M/\hbar^{1/4}$ to cure the problem of unitarity violation at frequencies above $m_{\mbox{\tiny{UV}}}\sim M/\hbar^{1/4}$. We expect that the new physics can renormalize the vacuum energy. Dimensional analysis selects a unique form:
\bea
\Delta\Lambda\sim\hbar\,m_{\rm new}^4
\eea

Of course we know that this theory is UV completed by QED (modulo the Landau pole) with the introduction of the electron with frequency 
\beq
m_{\rm new}=m_{\rm e}\sim\sqrt{\alpha}\,M/\hbar^{1/4}\sim\sqrt{\alpha}\,m_{\mbox{\tiny{UV}}} 
\eeq
which is parameterically smaller than the cutoff $m_{\mbox{\tiny{UV}}}$. Indeed the electron introduces a renormalization to the vacuum energy of the form $\Delta\Lambda \sim \hbar\,m_{\rm e}^4$, in accord with the above mentioned expectation. 

Hence even though the action in eq.~(\ref{EM}) does not by itself generate a large renormalization in vacuum energy, there is a large contribution introduced by the physics associated with its UV completion. This leads to a vacuum energy renormalization that is $\sim32$ orders of magnitude larger than the observed value. Hence a solution to this cosmological constant problem is to invent a mechanism by which $\Lambda\lll \hbar\,m_{\rm e}^4$ in a natural way in the theory with electrons.

\subsection{Pure Gravity}\label{Grav}

Let us now consider the case of pure Einstein gravity with standard action
\beq
S=\int\! d^4x\sqrt{-g}\left[-\Lambda+\mpl^2\mathcal{R}+\ldots\right]
\eeq
where $\mpl^2\equiv1/(16\pi G)$. Here we indicate a tower of higher derivative corrections by the dots. Let us expand around flat space by writing the metric as
\beq
g_{\mu\nu}=\eta_{\mu\nu}+h_{\mu\nu}/\mpl
\eeq
Focussing on the curvature term, this leads to an action that is schematically given by
\beq
S_R \sim \int\!d^4x\left[(\partial h)^2+h(\partial h)^2/\mpl+h^2(\partial h)^2/\mpl^2+\ldots\right]
\eeq
This theory leads to a large quantum correction to the bare vacuum energy of the form $\Lambda\sim\hbar\,\muv^4$, as usual.
However, there is not a significant flow in the {\em renormalized} vacuum energy, similar to the case of massless photons. Of course, it is also non-renormalizable and requires a UV completion. The theory requires new physics at a scale $m_{\rm new}<1/\sqrt{\hbar\,G}$ (the Planck frequency). Unlike the case of the interacting photons, here we do not know the form of the new physics to unitarize graviton scattering; it may involve extra dimensions, supersymmetry, strings, and/or other possibilities. 

In any case, we expect that the new physics will introduce contributions to shift the vacuum energy. In this case we can have at least two scales: (i) the scale that sets the new physics $m_{\rm new}$ (which might be many new scales), and (ii) (assuming the new physics permits a four-dimensional description in some regime) we also have Newton's constant $G$. This permits the following tower of dimensionally correct possibilities in increasing powers of $\hbar$:
\beq
\Delta\Lambda \sim \mpl^2m_{\rm new}^2,\,\hbar \, m_{\rm new}^4,\,\hbar^2\,m_{\rm new}^6/\mpl^2,\ldots
\label{rhograv}\eeq
The first term involves no powers of $\hbar$; it is a {\em classical} contribution from possible phase transitions, etc. It may or may not arise, depending on how the new physics interacts with gravity. The second term is the leading (1-loop) quantum contribution to the renormalization of $\Lambda$. Assuming new physics enters below the Planck scale, the higher order terms will be sub-dominant. The challenge to solve this cosmological constant problem is to find a mechanism in which $\Lambda\lll \mpl^2 m_{\rm new}^2,\,\hbar \, \muv^4,\ldots$. It would be some significant progress to have a dynamical mechanism wherein $\Lambda$ is naturally much less than the leading few terms, say, even if it is not smaller than the higher order terms. 

On the other hand, the absolute worst case scenario is to have a mechanism in which $\Lambda\sim \mpl^2m_{\rm new}^2$ or $\Lambda\gtrsim\hbar\,\muv^4$; this would not represent any progress at all. One might think this was progress if $\Lambda$ was small in the Planck units, due to $m_{\rm new}$ and/or $\muv$ being small, but this is not the real cosmological constant problem. If no Planck scale energies are permitted in the theory, then comparing to the Planck scale is irrelevant. The problem is to be small in terms of the energy scales that arise from the particle sector and in terms of the fundamental cutoff. Since (i) we know that even conventional Standard Model particle physics gives masses $m_{\rm new}\sim m_{t}\sim m_H\sim 100$\,GeV (which is ``new" physics from the low energy pure gravity point of view, even though these degrees of freedom may not be relevant to unitarizing graviton scattering), (ii) it is plausible that there are many heavier particles, such as at some extra dimension ``radion" scale, the GUT scale $m_{\rm new}\sim 10^{16}$\,GeV, and perhaps even heavier particles still, and (iii) the fundamental cutoff $\muv$ must be correspondingly even larger, if one was to obtain $\Lambda\sim \mpl^2m_{\rm new}^2$, or $\Lambda\gtrsim\hbar\,\muv^4$, it would be catastrophically large. Indeed one would anticipate that any purported dynamical solution at least achieves $\Lambda\lll \mpl^2m_{\rm new}^2,\,\hbar \, \muv^4$.

Shortly we will show that in a class of compactification models, $\Lambda\sim \mpl^2m_{\rm new}^2$ and $\Lambda\gtrsim\hbar \,\muv^4$ is generically obtained, which is indeed the worst case scenario and is therefore not a dynamical solution of the real problem.

\section{Compactification Models}\label{Comp}

Consider the following $D$-dimensional action involving gravity and a collection of $p$-form field strengths
\beq
S = \int\! d^{D}\!x\sqrt{-g_D}\left[-\Lambda_D+\mpls^{D-2} \mathcal{R}_D-\sum_p |F_p|^2\right]
\label{action}\eeq
where we have also included a higher-dimensional cosmological constant $\Lambda_D$ and the fundamental Planck mass is $\mpls^{D-2}\equiv 1/(16\pi G_D)$. We now illustrate how the four-dimensional theory emerges by compactification.

Here we consider a product manifold in the form $\mathbb{R}^4 \otimes {\cal M}^N $ where $\cal M$ is a $d$-dimensional manifold with a metric $h_{ij}$. This means we are considering $\exx\equiv N\,\ex$ extra dimensions.
We write the metric in the form 
\bea
	&&ds^2= e^{- d(\Psi(x)-\Psi_0)} g_{\mu \nu}(x) dx^\mu d x^\nu \nonumber\\
&&\,\,\,\,\,\,\,\,\,\,\,\,\,\,\,	+ \sum_a e^{2(\psi_a(x) - \psi_{0a})}h_{(a)ij} dy_{(a)}^{i}dy_{(a)}^j.
\label{metric}\eea
where $a$ is an index that runs from $a=1,\ldots,N$ over each of the internal manifolds of dimension $d$.
We have considered the simple case in which the higher-dimensional metric decomposes into a four-dimensional piece and a compact space with radion moduli $\rad_a=\rad_a(x)$ that only depends on the large dimensions. 
Here $\mu,\nu \in\{0,1,2,3\}$ and $x$ is the large dimension co-ordinates, 
while $i,j\in\{1,\ldots,\ex\}$. 
In the first term we have, without loss of generality, pulled out a factor of $e^{-\ex(\Psi(x)-\Psi_0)}$ so that the four-dimensional action is immediately in the Einstein frame, where 
\bel{Psi}
	\Psi(x)=\sum_a \psi_a(x), \qquad \Psi_0=\sum_a \psi_{0a}~.
\ee
and $\psi_{0a}$ is the value of $\psi_a$ at its stabilized value from compactification.


When integrating the fluxes over the compact space, we assume the fluxes are only functions of the compact co-ordinates, apart from an overall volume dependence. For the flux wrapping around the compact space labelled $(a)$, we have
\beq
\sum_p|F_p|^2 = \sum_p \mathcal{F}_{p(a)}(y) e^{-2p (\rad_a(x)-\rad_{a0})}
\eeq

We can now integrate over the compact space. This leads to 3 quantities that characterize its structure, namely
\bea
&& \mathcal{V}_{(a)} \equiv \int d^\ex y_{(a)}\sqrt{h_{(a)}}\\
&& C_{(a)} \equiv \int d^\ex y_{(a)}\sqrt{h_{(a)}}\,\mathcal{R}_{(a)}\label{Cdef}\\
&& f_{p(a)} \equiv \int d^\ex y_{(a)}\sqrt{h_{(a)}}\,\mathcal{F}_{p(a)}
\eea
where the first quantity $\mathcal{V}_{(a)}$ is the volume of each compact space, with total volume
\beq
\mathcal{V}=\prod_a \mathcal{V}_{(a)}, 
\eeq
the second quantity  $C_{(a)}$ is a volume integral over the compact space Ricci scalar $\mathcal{R}_{(a)}$, and the third quantity $f_{p(a)}$ is a volume integral over the flux number that threads the compact space. Note that each of these 3 quantities are constant; independent of space and time.

Dropping boundary terms, we obtain the following action in 4 dimensions
\bel{action2}
	S = \int d^4x \sqrt{-g_{(4)}} (-V_4+K_E)~,
\ee
where $K_E$ and $V_4$ are the kinetic terms and the four-dimensional potential term given by
\beal{KinetricPotential}
	V_4&=&e^{-d (\Psi-\Psi_0)}\Big{[}V_\Lambda  - \sum_a V_{C(a)}  e^{ - 2 (\psi_a-\psi_{a0})} \nn
	 &&\,\,\,\,\,\,\,\,\,\,\,\,\,\,\,\,\,\,\,\,\,\,\,\,\,\,\,\,+ \sum_{a,p} V_{p(a)} e^{-2p (\psi_a- \psi_{a0})}\Big{]},\label{V4}\\
	 	K_E&=&\mpl^2 \[{\mathcal R}  - \frac {d^2}2  \left( \nabla \Psi \right)^2 -  \sum_a d  \( \nabla \psi_a \)^2 \],
\eea
where 
\bel{defs}
	V_{\Lambda}\equiv\Lambda_D\,\CalV, \,\,\,\, V_{C(a)}\equiv {C_{(a)} \mpl^2\over\mathcal{V}_{(a)}},\,\,\,\, V_{p(a)}\equiv \frac{f_{p(a)} \CalV}{\mathcal{V}_{(a)}},
\ee
Here the four-dimensional Planck mass $\mpl$ is related to the fundamental Planck mass $\mpls$ by
\beq
\mpl^2=\mathcal{V}\,\mpls^{D-2}
\label{mpmps}\eeq

The above action involves scalar fields $\psi_a$ that are not canonically normalized. We can switch to a new set of fields $\phi_a$ that are canonically normalized by defining
\bel{diagonalize}
	\phi_a\equiv \sqrt{2d} \( \psi_a -\psi_{0a} + b(\Psi-\Psi_0)\)\mpl,
\ee
where
\beq
b\equiv\frac{-2+\sqrt{4 + 2 N d}}{2N}.
\eeq
which is defined such that the minima of the potential is at $\phi_a=0$.
In terms of these canonically normalized fields the action takes on the canonical form
\beq
S=\int d^4x\sqrt{-g}\left[-V(\phi_a)+\mpl^2\mathcal{R}-\sum_a{1\over2}(\partial\phi_a)^2\right]
\eeq
The potential function $V$ for the canonically normalized field is simply $V=V_4$, but expressed in different variables.

For a positively curved compact space, $C_{(a)}>0$, or for a negative higher-dimensional cosmological constant, $\Lambda_D<0$, one or both of the first two terms of $V$ is negative and it can compete with the other positive flux terms to lead to a stable vacuum solution, either AdS, Minkowski, or dS, depending on parameters. 

In the simplest versions of these models, with only one internal manifold $N=1$ and $\Lambda_D>0$ there can be dS vacua, but no accumulation of vacua as $\Lambda\to 0^+$ in the large flux limit. On the other hand, for $\Lambda_D\leq 0$ there is an accumulation of vacua with $\Lambda\to 0^-$ in the large flux limit. 

More interestingly, for $N\geq 2$ an accumulation point can arise for dS vacua as $\Lambda\to 0^+$ (as well as a much more dominant accumulation of vacua as $\Lambda\to 0^-$). This was shown in Ref.~\cite{Brown:2013fba} in the case where the internal manifolds are spheres.
This appears to be a beautiful solution of the cosmological constant problem. 

In the next section we will describe a general property of their solutions regarding moduli masses.

\section{Mass Scales}\label{Mass}

\subsection{Moduli Scale}\label{Moduli}

In this class of compactifications, and even for more general classes, the potential energy is a sum of terms involving exponentials of the radion fields $\phi_a$, as seen in eq.~(\ref{V4}). In a general fashion, we can write the potential as
\beq
V = \sum_q V_q\,\exp\left(- \sum_a \beta^{qa}\,\phi_a/\mpl\right)
\label{potsimp}\eeq
Comparing to (\ref{V4}) it is straightforward to read off the value of $V_q$ and $\beta^{qa}$ in this particular class of models. What is important to note is that the coefficients in the exponents $\beta^{qa}$ are $\mathcal{O}(1)$.
The mass of the moduli is given by the eigenvalues of the Hessian matrix of the potential at the minimum $\phi_a=0$. This is
\beq
H_{ab} = {\partial^2 V\over\partial\phi_a\partial\phi_b}\Bigg{|}_{\phi=0} = \sum_q {\beta^{aq}\beta^{bq}\over \mpl^2}\,V_q
\label{mphi}\eeq
Notice the eigenvalues cannot be much larger than the elements of $H_{ab}$, and so we will estimate the typical masses $m_a$ to be of order the typical values of $H_{ab}$. 
On the other hand, the cosmological constant is given by
\beq
\Lambda=V\big{|}_{\phi=0}=\sum_q V_q
\eeq 
Then assuming the potential does not have any accidental cancellations at its minima $\Lambda=V\big{|}_{\phi=0}$, and recalling that $\beta^{aq}=\mathcal{O}(1)$, we can express $\Lambda$ in terms of $m_a^2$ by comparing (\ref{potsimp}) to (\ref{mphi}), giving
\beq
\Lambda \sim \mpl^2\,m_a^2
\eeq
Then using the Friedmann equation, we have $m_a\sim H$ quite generally, where $H$ is the Hubble parameter. So we see that generically the moduli mass is related to the Hubble scale, which is reasonable on dimensional grounds. 

Typically, most vacua are AdS. One can restrict attention to only dS vacua (as was the case in Refs.~\cite{Brown:2013fba,Brown:2014sba}). These dS vacua require a fine-tuning to achieve a special cancellation in the potential  between the large $V_\Lambda$ contribution and a curvature contribution, leaving $\Lambda$ especially small (see Section \ref{nongeneric} for more details). However, one can show that a typical dS vacuum has a corresponding special cancellation in the potential, leaving $\Lambda\sim \mpl^2\,m_a^2$ still valid for light moduli (although some moduli can be heavier).

In more complicated models, even if one were to obtain $\Lambda\ll \mpl^2\,m_a^2$, there is no evidence that $\Lambda\ll\hbar \, m_a^4$ can be obtained within this framework.



Hence from the point of view of the renormalization group flow of $\Lambda$, we see that these models do not produce a $\Lambda$ that is smaller than the typical expectation from renormalization. In the next section, we will study whether $\Lambda$ is much smaller than the estimates based on a sharp cutoff.

\subsection{Fundamental Scale}\label{Fund}

\subsubsection{AdS Vacua (generic)}

In this class of models, the higher-dimensional cosmological constant $\Lambda_D$ plays a very important role. In most compactifications, it provides an $\mathcal{O}(1)$ contribution to the four-dimensional vacuum energy. As shown in \cite{Brown:2014sba} most of these vacua are AdS.
For these vacua the four-dimensional vacuum energy is
\beq
\Lambda\sim -V_{\Lambda}=-\Lambda_D\,\mathcal{V}
\eeq
Now suppose we parameterize the higher-dimensional cosmological constant as
\beq
\Lambda_D = \lambda_D\,\mpls^{D}
\eeq
So if $\lambda_D$ is chosen to be $\lambda_D=\mathcal{O}(1)$ (in units of $\hbar$), then we have a ``naturally" large value for the higher-dimensional cosmological constant, according to the ``sharp cutoff analysis" of Section \ref{Sharp}.

Now, on the one hand, we can eliminate $\mpls$ in the expression for $\Lambda$, by using eq.~(\ref{mpmps}) leading to
\beq
\Lambda \sim - \lambda_D\,\mpl^4\left(\lpl^{\exx}\over\mathcal{V}\right)^{\!2/(\exx+2)}
\label{LammpV}\eeq
where $\exx = N\,\ex$.
Since these are large volume compactifications, in the large flux limit, we see that $\Lambda\to0^{-}$, which appears to solve the problem of why the cosmological constant is small (although this is negative). 
On the other hand, we can eliminate $\mathcal{V}$ and express $\Lambda$ in terms of $\mpls$ leading to 
\beq
\Lambda \sim - \lambda_D\,\mpl^2\mpls^2
\eeq
For $\lambda_D$ not too small, this $\Lambda$ is {\em much larger} than even the ``natural" value of $\sim \mpls^4$ (in units of $\hbar$), since $M_*\ll\mpl$ in the large volume limit. Hence this clearly does not address the cosmological constant problem. We see that the only reason $\Lambda\to 0^{-}$ is because $\mpls\to 0$ which removes all high energy physics trivially.

\subsubsection{dS Vacua (non-generic)}\label{nongeneric}

Alternatively, one can introduce very special choices of flux parameters, so as to fine-tune away such huge contributions to the vacuum energy, and allow for dS vacua. To do this, we need a hierarchy among the internal radii, which leads to a hierarchy among the curvature contributions $V_{C(a)}$. There are two important possibilities:
\begin{enumerate}[{(i)}]
\item $ V_{C(1)} \ll V_{C(2)}\sim \cdots \sim V_{C(N)} $
\item $ V_{C(1)} \gg V_{C(2)}\sim \cdots \sim V_{C(N)} $
\end{enumerate}

In the first case (i), the curvature contribution $\sum_{a=2}^N V_{C(a)}$ is tuned to cancel against the vacuum energy contribution $V_\Lambda$.
The residual four dimensional vacuum energy can be estimated by the residual curvature contribution
\beq
\Lambda\sim V_{C(1)} 
\eeq
From eq.~(\ref{Cdef}), the curvature parameter $C_{(1)}$ is roughly $C_{(1)}\sim\mathcal{V}_{(1)}^{1-2/\ex}$. Expressing $\Lambda$ in terms of $\mpl$ and $\mathcal{V}_{(1)}$ gives
\beq
\Lambda\sim\mpl^4\left(\lpl^{\ex}\over\mathcal{V}_{(1)}\right)^{\!2/\ex}
\label{curv4a}
\eeq
which shows that indeed the vacuum energy is even smaller than previously in eq.~(\ref{LammpV}), for large volumes $\mathcal{V}_{(1)}$. This is the result of fine-tuning the leading contributions to vanish, and allows $\Lambda\to 0^{+}$ more rapidly, which appears to solve the problem of why the cosmological constant is small and positive. However, we can again eliminate $\mathcal{V}_{(1)}$ and express the result in terms of $\mpls$, to find
\beq
\Lambda \sim {\mpls^4\over\lambda_D^{N-1}}\left(\mpl\over\mpls\right)^{\!2-{4\over\ex}}
\eeq
We note that for curvature to be present, we obviously need $\ex\geq 2$. So $\Lambda\gtrsim \mpls^4$ (in units of $\hbar$)
and is bounded by $\Lambda\lesssim \mpls^2\mpl^2$ for high $d$.

In the second case (ii), the curvature contribution $V_{C(1)}$ is tuned to cancel against the vacuum energy contribution $V_\Lambda$.
The residual four dimensional vacuum energy can be estimated by the residual curvature contribution
\beq
\Lambda\sim V_{C(a)},\,\,\,\,\mbox{where}\,\,\,a=2,\ldots,N 
\eeq
So we can write an expression similar to eq.~(\ref{curv4a}) as
\beq
\Lambda\sim\mpl^4\left(\lpl^{\ex}\over\mathcal{V}_{(a)}\right)^{\!2/\ex}
\label{curv4b}
\eeq
using the fact that $\mathcal{V}_{(a)}$ are all similar for $a=2,\ldots,N$ in this case. This again says that the vacuum energy tends to $\Lambda\to 0^{+}$ in the large volume limit. We now eliminate the volume dependence to express the result in terms of $M_*$, as we did in case (i), to find
\beq
\Lambda \sim {\mpls^4\over\lambda_D^{1/(N-1)}}\left(\mpl\over\mpls\right)^{\!2-{4\over\ex(N-1)}}
\eeq
Since we need $d\geq2$ for curvature to be present and $N\geq2$ for this cancellation to take place, we again have $\Lambda\gtrsim \mpls^4$ (in units of $\hbar$) and is bounded by $\Lambda\lesssim \mpls^2\mpl^2$ for high $d$ or $N$. 

So in both cases (i) and (ii) we see that despite having fine-tuned away the leading term to produce dS vacua, the resulting cosmological constant is still not smaller than the estimate based on a simple cutoff.

\section{Conclusions}\label{Disc}

Hence even though there are interesting compactification models \cite{Brown:2013fba,Brown:2014sba} in which the four-dimensional cosmological constant has an accumulation point as $\Lambda\to 0$, it does so only in so far as the mass scales of fundamental physics $m_{\rm new},\,\muv\to 0$. Generally in these models, $\Lambda$ scales as some power of the product of the mass and Planck mass appearing in the four-dimensional theory, or alternatively, as a power of the fundamental Planck scale. This is essentially the worst case scenario from both the renormalization group point of view, and also from the sharp cutoff point of view. This does not address the real cosmological constant problem, wherein we need to explain how $\Lambda$ is incredibly small, despite the presence of high scales of physics.

A general way to see the problem is the following: From the low energy four-dimensional point of view, it cannot matter that there are extra dimensions, or otherwise, in the UV.  Effective field theory says that these effects cannot naturally reach down and remove the already large contributions to vacuum energy from known Standard Model physics. In the above toy models, an attempt to do this comes from having the new ``UV" physics scale simply inserted at fantastically low energies, which misses the real problem.

Returning to the structure of eqs.~(\ref{potsimp}) we see that the only possibility would be to allow the potential $V$ to have various {\em accidental} cancellations at its minimum $\Lambda$, while maintaining large mass scales $m_\phi,\,\mpls$. This could be conceivable in some landscape framework with many fluxes and an exponentially large number of vacua, and is a conceivable solution \cite{Sakharov:1984ir,Bousso:2000xa}. Other directions to address the problem could involve radical alterations to local quantum field theory.

\begin{center}
{\bf Acknowledgments}
\end{center}
We would like to thank Alex Vilenkin and Erick Weinberg for helpful discussions.
This work is supported by the U.S. Department of Energy under cooperative research agreement Contract Number DE-FG02-05ER41360. AM is supported by a grant from National Science Foundation (grant PHY-1213888).


\end{document}